# Skyrmion-Magnetic Tunnel Junction Synapse with Mixed Synaptic Plasticity for Neuromorphic Computing

*Aijaz H. Lone[1], Arnab Ganguly[2], Selma Amara[1], Gobind Das[2] and H. Fariborzi[1]*

[1]*Computer, Electrical and Mathematical Science and Engineering Division, King Abdullah University of Science and Technology, Thuwal, Saudi Arabia.*

[2]*Department of Physics, Khalifa University, Abu Dhabi 12788, United Arab Emirates.*

Magnetic skyrmion-based data storage and unconventional computing devices have gained increasing attention due to their topological protection, small size, and low driving current. However, skyrmion creation, deletion, and motion are still being studied. In this study, we propose a skyrmion-based neuromorphic magnetic tunnel junction (MTJ) device with both long- and short-term plasticity (LTP and STP) (mixed synaptic plasticity). We showed that plasticity could be controlled by magnetic field, spin-orbit torque (SOT), and the voltage-controlled magnetic anisotropy (VCMA) switching mechanism. LTP depends on the skyrmion density and is manipulated by the SOT and magnetic field while STP is controlled by the VCMA. The LTP property of the device was utilized for static image recognition. By incorporating the STP feature, the device gained additional temporal filtering ability and could adapt to a dynamic environment. The skyrmions were conserved and confined to a nanotrack to minimize the skyrmion nucleation energy. The synapse device was trained and tested for emulating a deep neural network. We observed that when the skyrmion density was increased, the inference accuracy improved: 90% accuracy was achieved by the system at the highest density. We further demonstrated the dynamic environment learning and inference capabilities of the proposed device.

*Index Terms*—Skyrmions, magnetic tunnel junction (MTJ), long term plasticity (LTP), short term plasticity (STP), dynamic environment learning, and neuromorphic computing

## I. INTRODUCTION

Magnetic skyrmions are topologically protected swirling structures induced by chiral interactions in non-centrosymmetric magnetic compounds or thin films with broken inversion symmetry [1]. The Dzyaloshinskii–Moriya interaction (DMI)—the chiral antisymmetric exchange interaction responsible for the formation of these textures—originates due to strong spin-orbit coupling at the heavy metal/ferromagnetic interface (HM/FM) with broken inversion symmetry [2]. These magnetic textures emerge out of the competition between different energy terms. The exchange and anisotropy terms prefer the parallel alignment of spins, whereas the DMI and dipolar energy terms prefer the non-collinear alignment of spins[3]. In an asymmetric ferromagnetic multilayer system, such as Pt/Co/Ta [4] and Pt/CoFeB)/MgO [5], the DMI is induced by the high interfacial spin-orbit coupling resulting from symmetric breaking. Since DMI and anisotropy are material properties and geometry-dependent, to stabilize these skyrmions and define a particular chirality, combinations of different HM/FM structures are being investigated [6]. The spintronic devices based on these textures promise increased density and energy-efficient data storage due to their small nanometric size and topological protection [7]. These textures can be driven by very low depinning current densities [8] and show scalability down to 1 nm [9]. These superior properties of magnetic skyrmions reveal bright prospects for efficient data storage and computing [10]. In particular, skyrmion devices show great potential for unconventional computing, such as neuromorphic computing [11] and reversible computing [12]. Neuromorphic computing has been inspired by the brain's performance and energy efficiency [13] and involves neuro-mimetic devices, such as neurons, responsible for computing. The synapses store information in terms of weight. Spintronics devices, especially the magnetic tunnel junction (MTJ), have found wide applicability in neuromorphic computing [14][15]. In recent years, different neuromorphic computing systems with MTJ devices based on skyrmions have been proposed, such as skyrmion neurons[16][13] and skyrmion synapses [11][17]. Furthermore, the electric-field control of spintronic devices has been receiving increasing attention in memory and logic applications, as they provide an efficient way to improve the data storage density [18][19] and reduce the switching energy [20]. However, the vital challenge associated with the application of skyrmions for storage and computing, for both conventional and unconventional computing, is the controlled motion and readability of skyrmions [21].

This study proposes a skyrmion-based neuromorphic MTJ (Ta/CoFeB/MgO/CoFeB) device with both long- and short-term plasticity (LTP and STP) (mixed synaptic plasticity). We confirmed plasticity control by magnetic field, spin-orbit torque (SOT), and the voltage-controlled magnetic anisotropy (VCMA) switching mechanism. LTP depends on the skyrmion density, which is manipulated by the SOT and magnetic field, and STP is controlled by the VCMA. The LTP property of the device was utilized for static image recognition. By introducing the STP feature, the device gained additional temporal filtering ability and could adapt to a dynamic environment. The skyrmions were conserved and confined to a nanotrack to minimize the skyrmion nucleation energy. The synapse device was trained and tested for the system-level emulation of a deep neural network based on the Canadian Institute for Advanced Research (CIFAR-10) data set. We observed that when the skyrmion density was increased, the inference accuracy improved: 90% accuracy was achieved by the system at the highest density.

The rest of the paper is organized as follows. Sample preparation and characterization will be discussed in section II. In section III, we will explore device modeling using micromagnetics and the non-Equilibrium Green's function (NEGF) formalism. In section IV, we will outline our findings and present a discussion on skyrmion size and density dependence on different energy terms. Section V introduces our proposed neuro-mimetic devices (skyrmion synapses and neurons) and their electrical manipulation for system-level applications. The concluding remarks will be presented in the last section, which will be followed by a discussion on the future prospects of our work, such as the potential reservoir computing.

## II. FABRICATION AND CHARACTERIZATION

A thin-film multilayer of Ta (5 nm)/CoFeB (1.04 nm)/MgO (2 nm)/Ta2(nm), as shown in Fig. 1, was deposited on thermally oxidized Si substrates using Singulus DC/RF magnetron sputtering. The CoFeB thickness was a curtail parameter providing suitable anisotropy for generating high-density skyrmions. The sputtering conditions were carefully optimized for obtaining perpendicular magnetic anisotropy (PMA). The sample was subjected to post-deposition annealing at 250° C for 30 min to further enhance its PMA. The experiment was performed using magneto-optical Kerr effect (MOKE) microscopy in polar geometry. Differential Kerr imaging was performed to observe the magnetic domains and eliminate the contribution of any non-magnetic intensities. The square pulses of the magnetic field were simultaneously applied, both in-plane and out-of-plane, to the sample using two independent electromagnets. The sample exhibited a labyrinth domain structure in the absence of any magnetic field. Saturated magnetization was first observed in one perpendicular direction by a sufficiently large out-of-plane field. A reference p-MOKE image was captured in this state. In the next step, $H_Z$ was reduced to the desired value accompanied by an in-plane field $H_X$, as required. A second p-MOKE image was captured in this state. The magnetic image of the final state was obtained by the differential image with respect to its reference image

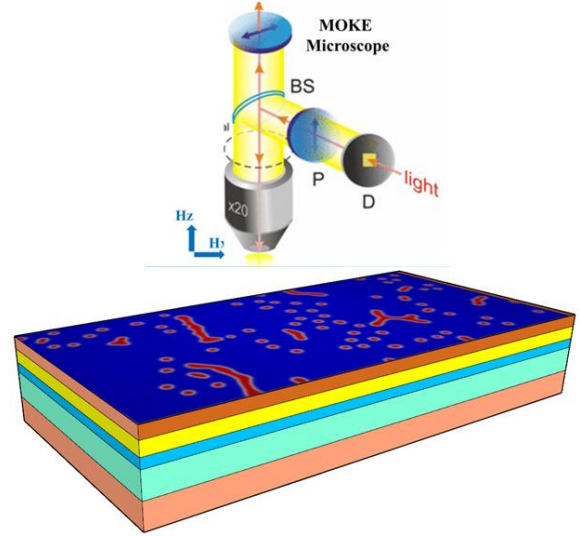

Figure 1. (a) Schematic of the fabricated sample for skyrmion detection. The stack comprises Ta(5nm) as the heavy metal. At the interface with CoFeB(1.04nm), DMI was generated, with MgO acting as the inversion symmetry breaking layer and Ta(2nm) as the capping layer. (b) Schematic of the stack under MOKE measurement.

## III. DEVICE MODELING

Magnetic skyrmions are defined by their topological number or skyrmion number, Q, and are calculated by [22]

$$Q = \frac{1}{4\pi} \int \int \boldsymbol{m} \cdot \left(\frac{\partial \boldsymbol{m}}{\partial x} \times \frac{\partial \boldsymbol{m}}{\partial y}\right) dx dy \quad (1)$$

The spins are projected on the x-y plane. The normalized magnetization vector ($\boldsymbol{m}$) can be determined by the radial function $\theta$, vorticity $Q_v$, and helicity $Q_h$.

where the helicity number is related to the skyrmion number through the following expression [10]:

$$Q = \frac{Q_v}{2}\left[\lim_{r \to \infty} \cos(\theta(r)) - \cos(\theta(0))\right] \quad (3)$$

The micromagnetic simulations were carried out using MuMax, incorporating Landau–Lipschitz–Gilbert (LLG) equation as the basic magnetization dynamics computing unit [23]. The LLG describes magnetization evolution as follows:

$$\frac{d\hat{m}}{dt} = \frac{-\gamma}{1+\alpha^2}\left[\boldsymbol{m} \times \boldsymbol{H}_{eff} + \boldsymbol{m} \times (\boldsymbol{m} \times \boldsymbol{H}_{eff})\right] \quad (4)$$

where $\boldsymbol{m}$ is the normalized magnetization vector, $\gamma$ is the gyromagnetic ratio, $\alpha$ is the Gilbert damping coefficient and

$$\boldsymbol{H}_{eff} = \frac{-1}{\mu_0 M_S} \frac{\delta E}{\delta \boldsymbol{m}} \quad (5)$$

Where *Heff* is the effective magnetic field around which magnetization process. The total magnetic energy of the free layer includes the exchange energy, Zeeman energy, uniaxial anisotropy energy, demagnetization energy, and DMI energy [24][25].

$$E(\boldsymbol{m}) = \int_V [A(\nabla \boldsymbol{m})^2 - \mu_0 \boldsymbol{m} \cdot H_{ext} - \frac{\mu_0}{2} \boldsymbol{m} \cdot H_d - K_u(\hat{u} \cdot \boldsymbol{m}) + \varepsilon_{DM}]dv \quad (6)$$

where A is the exchange stiffness, $\mu_0$ is the permeability, Ku is the anisotropy energy density, $H_d$ is the demagnetization field, and $H_{ext}$ is the external field. The DMI energy density can be computed by

$$\varepsilon_{DM} = D[m_z(\nabla \cdot \boldsymbol{m}) - (\boldsymbol{m} \cdot \nabla) \cdot \boldsymbol{m}] \quad (7)$$

The SOT was added as the custom field term in MuMax [26].

$$\boldsymbol{\tau}_{SOT} = -\frac{\gamma}{1+\alpha^2}a_J[(1+\xi\alpha)\boldsymbol{m}\times(\boldsymbol{m}\times\boldsymbol{p})+(\xi-\alpha)(\boldsymbol{m}\times\boldsymbol{p})]$$

$$a_J = \left|\frac{\hbar}{2M_Se\mu_0}\frac{\theta_{SH}j}{d}\right| \quad \text{and} \quad \boldsymbol{p} = sign(\theta_{SH})\boldsymbol{j}\times\boldsymbol{n} \quad (8)$$

where $\theta_{SH}$ is the spin Hall coefficient of the material, $j$ is the current density, and d is the free layer thickness. The synapse resistance and neuron output voltage can be computed by using the NEGF formalism. We considered the magnetization profile of the free layer and fed it to our NEGF model, which computes the resistance of the MTJ device as follows [27][28]:

$$R_{syn} = \frac{V_{syn}}{I_{syn}} \quad (9)$$

The MTJ read current can be computed by

$$I_{syn} = trace\left\{\sum_{k_t} C_\sigma \frac{i}{\hbar}\begin{Bmatrix}H_{k,k+1}G^n_{k+1,k}\\-G^n_{k,k+1}H_{k+1,k}\end{Bmatrix}\right\} \quad (10)$$

where $H_k$ is the kth lattice site in device Hamiltonian, and $G^n_k$ is the electron correlation at the kth site that yields the electron density.

## IV. SKYRMION-MTJ SYNAPSE WITH MIXED PLASTICITY (LTP AND STP), CONTROLLED BY SOT AND VCMA

Fig. 2 shows our proposed skyrmion-MTJ synaptic device structure *(Ta(5nm)/CoFeB(1nm)/MgO(2nm)/CoFeB(2nm))*, based on skyrmion size and density manipulation. We divided the synapse into three regions: pre-synapse, active synapse, and post-synapse. The skyrmions in the free layer were driven toward the active MTJ region, where we read them in terms of change in magnetization in the active region. While driving skyrmions, it is important to ensure that they do not leave the active region before getting read by the MTJ due to their SOT and inertia'. To ensure this, we used a VCMA gate control, as shown in Fig. 2. The red strip on the right side of the active region indicates that a voltage pulse was applied. Further, the anisotropy in this strip was changed, causing an energy barrier for the skyrmions to move toward its right. We applied a positive current to drive the skyrmions from the pre-synaptic region into the active region. Upon increasing the number of pulses, the skyrmion density increased, which was read in terms of the overall change in magnetization in the active region. This magnetization was rendered as an input to the NEGF model to calculate the change in resistance (decrease) in the synapse causing synaptic potentiation. These resistance states tuned by SOT acted as weights for the neural network. Synaptic depression was realized by removing the VCMA gate pulse from the right strip and applying current in the x direction (same as for potentiation). We observed that the majority of skyrmions left the active region and accumulated in the post-synapse region, whereby the resistance of the active region increased. To again increase the weight, we reversed the SOT current polarity and turned on VCMA gate 1. The same procedure was followed as in the previous case. Thus, we potentiated and depressed the synapse weight using unipolar current. Fig. 6(b) depicts the equivalent circuit model involving the mixed plasticity (LTP+STP) synapse. The additional degree of tuning for plasticity made the device more adaptable to dynamic environments, qualifying this behavior as more relevant to the mammalian brain.

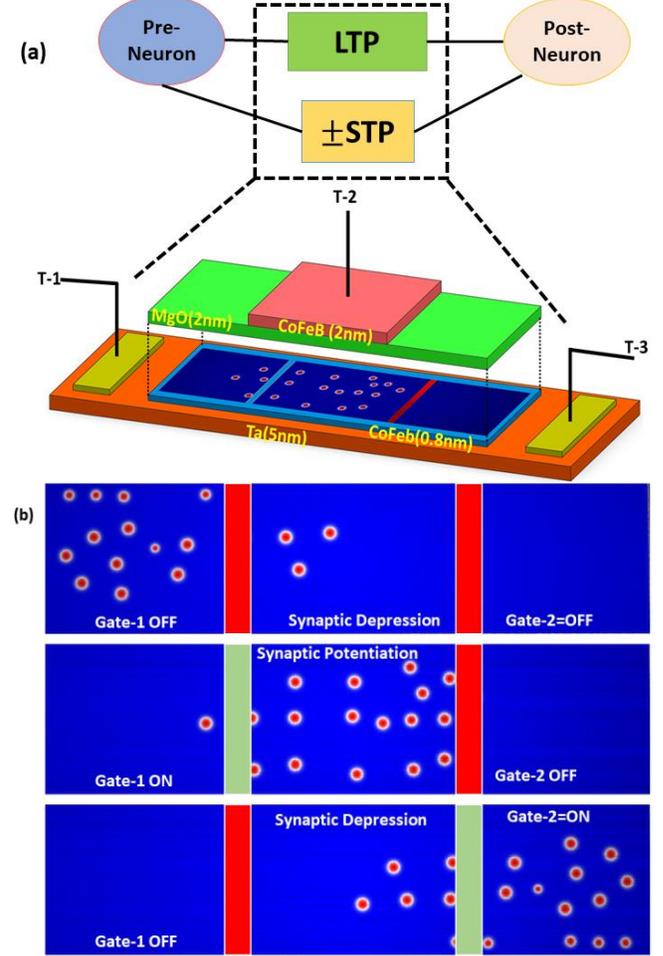

Fig. 2(a) SOT-driven and VCMA-controlled skyrmion-MTJ synapse device structure (Ta/CoFeB/MgO/CoFeB). A neural circuit model with pre- and post-neurons with skyrmion-MTJ synapse showing both LTP and STP (mixed plasticity). (b) Operating phases of the device for realizing synaptic potentiation and synaptic depression.

The complete operating description is shown in Fig. 6(c). Depending on the density of skyrmions in the two regions (pre-synapse or post-synapse), we set the current direction accordingly.

In Fig. 2(a), the light blue boundaries in the free layer indicate high anisotropy across the edges for confining these skyrmions to a fixed track. The confinement brings in the advantage of conservative computing. We could either nucleate and stabilize the fixed density of the skyrmions in a controlled manner at the beginning or we could expect the skyrmions to exist at zero field. Thereafter, we just tinkered with the skyrmion motion and size to realize the synaptic behavior. (Note: the driving current has to be below the annihilation current for reliable operation.) The device-circuit-system methodology for the system-level performance of the skyrmion-MTJ synaptic device is shown in Fig. 3(a). The skyrmion density from the MOKE measurements and SOTs were input to the MuMax simulator, which computed the variations in the magnetization profile due to skyrmion density and SOT in the free layer. This magnetization profile was fed to the NEGF (MATLAB) module to compute the synaptic resistance evolution and effective tunnel magnetoresistance (TMR) of the device. These resistance states and TMR were provided as input to the DNN+NeuroSim simulator, which evaluated the chip level performance of the proposed synapse device.

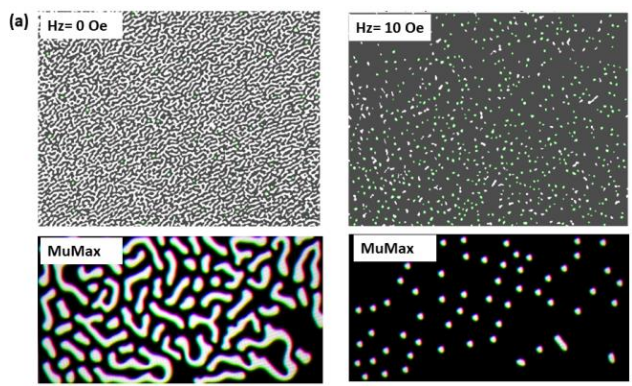

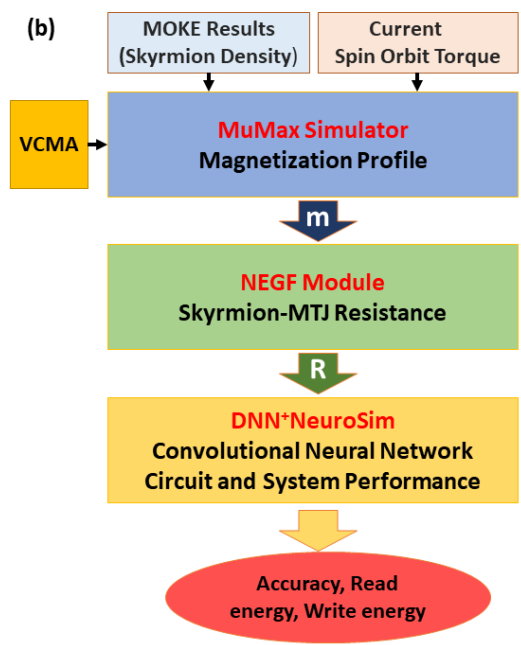

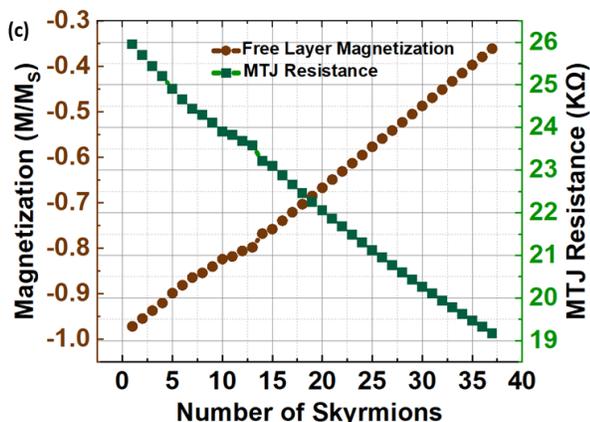

Fig. 3(a) MOKE and MuMax captures of the free layer showing the existence of skyrmions at the non-zero magnetic field (10 Oe). (b) The skyrmion density changed the free layer magnetization. Increasing the skyrmion number reduced the MTJ resistance (computed by the NEGF module).

Fig. 3(b) shows the magnetization change with voltage pulses during potentiation and depression. Compared to the synaptic behavior with SOT, we observed more non-linearity in the case of VCMA control, which could be attributed to the low relaxation time during VCMA operation. Fig. 3(c) shows the synaptic magnetization and resistance evolution with the SOT current pulses. An increase in the number of pulses increased the skyrmion density. Thus, the magnetization of the free layer changed and synaptic resistance decreased (potentiation). The current manipulated the skyrmion density causing long-term potentiation/depression (LTP/LTD) as the synapse weight remained non-volatile for a longer period.

We further used the VCMA in the active region for tuning the skyrmion size and density, causing short-term potentiation/depression in the same synapse, as shown in Figs. 4(a)–c. In mammalian brains, it is believed that the mixed type of synaptic learning, i.e., STP and LTP, is used. STP makes the system adaptable to dynamic environments. The relaxation time for skyrmion size is very short when anisotropy is changed momentarily. The skyrmion size either increases or decreases. However, when the voltage is removed, the skyrmion relaxes back to its original size. Fig. 4(a) shows the voltage-controlled STP for a fixed skyrmion density. Upon applying positive voltage pulses of increasing magnitude, the free layer anisotropy decreased by 1–5% depending on the magnitude of the VCMA. As the anisotropy increased, the skyrmion size also increased, resulting in a positive magnetization change. Likewise, when negative voltage pulses were applied, the free layer anisotropy increased, causing the skyrmion size to decrease, as shown in Fig. 4(a). We fed the magnetization variation into the NEGF module, which computed this modulation in terms of the synaptic resistance. This magnetization variation was represented in terms of short-term potentiation (STP) and short-term depression (STD). In Fig. 4(b), the STP at different skyrmion densities (but constant voltage) indicates that the overall change in magnetization increased as expected; however, with the increase in density, the skyrmion–skyrmion repulsion also increased, which decreased the skyrmion size, leading to the overall magnetization of the free layer. Regarding the lower skyrmion density (8 skyrmions), it was observed that the magnetization first jumped to a peak, decreased a little, and again increased with time until it stabilized at around the initially attained peak. However, for the 24 skyrmions, the magnetization jumped to the peak value of 9%, following which, due to the increased skyrmion–skyrmion interaction, the skyrmion size decreased. After 1 ns, the magnetization stabilized at 4%, which was about half for the 8 skyrmions. The synaptic STP for the three skyrmion densities (5, 15, and 25) is illustrated in Fig. 4(c). The magnetization of the active region decreased with an increase in voltage. We also observed an increase in STP. However, owing to the skyrmion–skyrmion interaction, the overall STP was almost the same for 5 and 15 skyrmions. The results in Figs .4(b) and 4(c) show that it is important to optimize the skyrmion density in the MTJ so that a large voltage-based STP is induced. Further, in comparison to LTP, we observed non-linear synaptic behavior for the VCMA-controlled synapse with a leaky component involved. This behavior was a clear depiction of mixed plasticity (LTP+STP). In Fig. 4(d), we increased the voltage from 0 V to 3.8 V in steps of 0.3 V and then reduced it from 3.8 V down to 0 V in the same number of steps. Regardless of the short relaxation time, we still observed the typical memristive behavior, as exhibited in the hysteresis loop in Fig. 4(d).
We formulated the mixed LTP and STP in the same skyrmion-MTJ device by

$$G_{t+1} = G_t^{LTP} \pm G_t^{STP} \qquad (11)$$

where $G_t^{LTP}$ is the fixed conductance/weight at time t due to the LTP characteristics of the device, and $G_t^{STP}$ is the dynamic STP conductance. Depending on the sign of the input spikes, the overall weight of the synapse could either increase or decrease. $G_t^{STP}$ is the function of skyrmion density and voltage spikes from the pre-synaptic neurons.

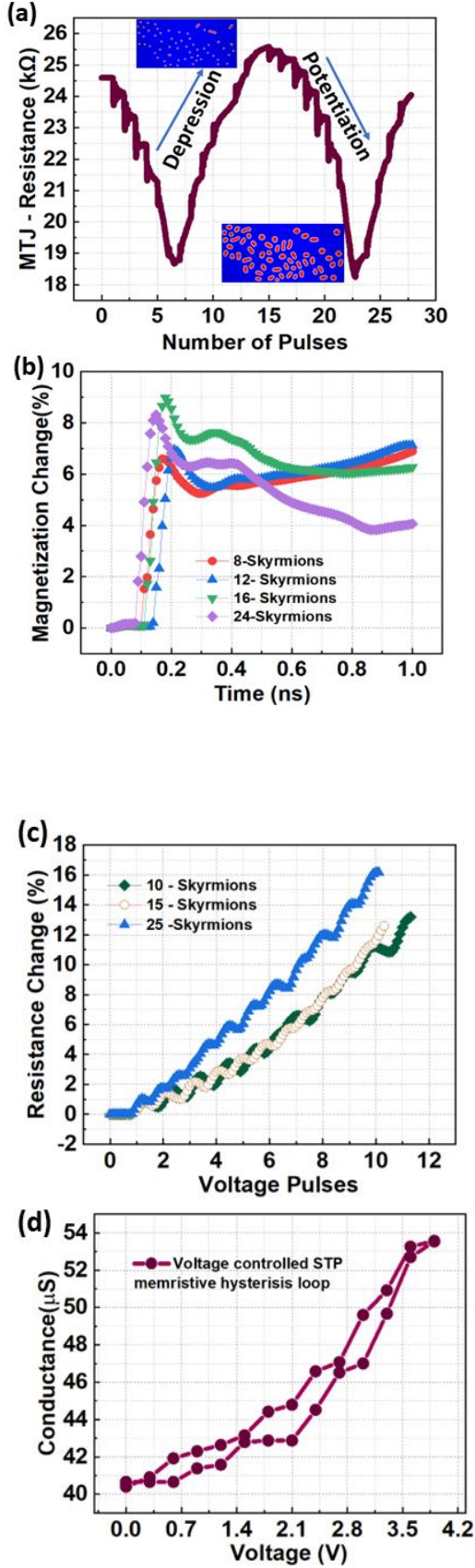

Fig. 4(a) Voltage-controlled short-term plasticity (short-term potentiation (STP) and short-term depression (STD)). (b) Short-term plasticity at different skyrmion densities but constant voltage. (c) Short-term resistance changes due to VCMA for increasing skyrmion density. (d) STP hysteresis showing typical memristive behavior.

$$G_t^{STP} = f(SKD, \sum_{i=1}^{N} V_i) \qquad (12)$$

## V. SKYRMION-MTJ SYNAPSE-BASED NEUROMORPHIC COMPUTING FOR IMAGE RECOGNITION

The circuit- and system-level neuromorphic computing performance, such as on-chip training and inference of the proposed skyrmion-MTJ synapse, was tested on the in-memory computing hardware accelerator architecture. We employed an integrated framework, DNN+NeuroSim [29], for this task. Figs. 5(a) and 5(b) show the neural network for on-chip training and inference based on the skyrmion-MTJ. As seen in Fig. 8(b), the synaptic array was in 2T-1R configuration. We connected the SOT terminals with the bit line, and the word/write line was connected to the gate of transistor 1. We adopted the default 8-Layer VGG-8 convolution neural network from the Canadian Institute for Advanced Research (CIFAR-10) data set for training and inference. The network comprises the first 1–6 convolution layers followed by two fully connected layers, as shown in Fig. 5(a) The details of the network are included in the supplementary. The skyrmion-MTJ synapse parameters used for the simulations are also outlined in the supplementary. The number of skyrmions determines the number of conductance levels and the on/off ratio of the synapse. Considering the increase in skyrmion–skyrmion repulsion with an increase in skyrmion density, we could not expect perfect TMR in the proposed geometry, with skyrmions as the source of information. Thus, depending on the skyrmion fill factor (area covered by the skyrmions/total area of the active synapse), we defined the effective TMR, which increased with the skyrmion density. For example, a 50% effective TMR may correspond to a 90% real TMR of the MTJ. We input the data on skyrmion density, manipulated by the magnetic field, current, and voltage (anisotropy), to the simulator for performance evaluation of the network. As expected, the accuracy increased with an increase in the number of skyrmions in the active region. The impact of increased density reflected in the overall magnetization change of the MTJ. Consequently, this increased the synaptic conductance, which increased the on/off ratio. The accuracy for 50% TMR after 40 epochs was 84%; for 70% TMR, corresponding to the on/off ratio of 1.7, the accuracy increased to 88%; and for 100% TMR (with on/off ratio=2), the accuracy was 89%, as shown in Fig. 5(c). Thus, a 70% TMR in the proposed synapse was good enough for classification. Considering the well-separated skyrmion motion during potentiation, the synaptic behavior exhibited very low non-linearity. However, as the density increased, we expected some non-linearity coming into the picture due to skyrmion–skyrmion interactions. Nevertheless, compared to other beyond-CMOS technologies, the non-linearity can be controlled with proper writing. By increasing the skyrmion density and improving the TMR of the MTJ, these skyrmion-MTJ devices can come close to achieving an accuracy of 93%, performing at par with software counterparts. The other circuit-level and chip-level performance metrics, such as read energy, write energy, synaptic core area, and memory utilization, are provided in the supplementary material. In Fig. 5(c), the second image on the top right-hand corner shows the weight distribution of the fully connected layers FC-1 (channel width=length=1, depth=8192, and kernel size=1024) after training. The bottom left- and right-hand corner images in Fig. 5(c) show the weight distribution before and after training for the fully connected layers FC-2 (channel width=length=1, depth=1024, and kernel size=10). The chip-level performance of the proposed device is included in the supplementary material.

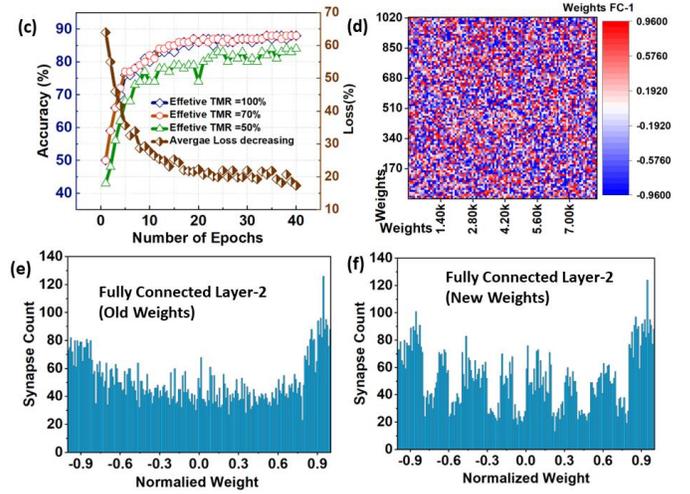

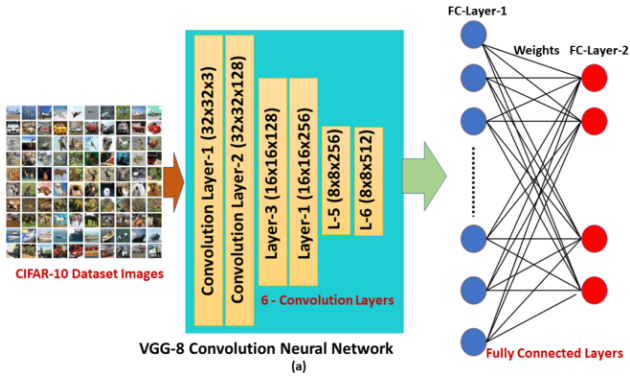

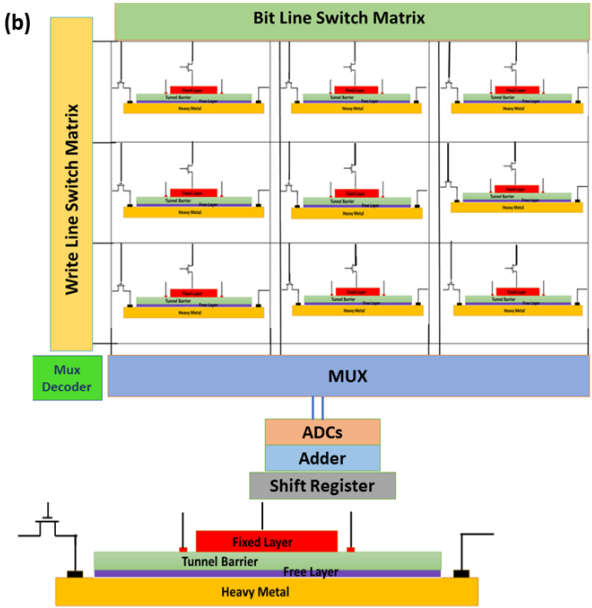

Fig. 5(a) VGG-8 convolution neural network implementation by employing the proposed skyrmion-MTJ synapse. The DNN was trained and tested on the CIFAR-10 data set. (b) Accuracy versus the number of epochs for varying effective TMR ratios, indicating that as the TMR increases (corresponding to an increase in skyrmion density), the accuracy increases, reducing network loss. (c) Weight distribution after learning by the fully connected layer 1 (d) Weight distribution of the fully connected layer 2 before training and (e) after training.

To demonstrate the dynamic environment learning capabilities of the proposed device, we prepared a 4x4 pixels-based digit recognition of handwritten digit 1, as shown in Fig. 6. The pixel values were mapped to the input neuron layer (16 neurons). These input neurons were connected to the three output neurons representing digits 1, 2, and 0. We set the synaptic weights G (S2, S6, S10, and S14) to the maximum conductance and the rest of the synapses to a minimum. This configuration corresponds to the static pattern 1. Corresponding to each blue pixel, 0.5 V was fed to the network. White pixels are represented by 0 V. These voltage signals input a current to the fixed resistor at the MTJ input. Based on the configuration, the highest conductance provided the maximum current. The lowest conductance was normally supplied with 0 V, leading to zero current. We considered a simple threshold neuron that shows a threshold value. If the input voltage at the MTJ was above VTH=3 V, the neuron generated a spike; otherwise, no spike was generated. As shown in Fig. 6(b) for the first 150 ns, we kept the pattern static and observed that the neuron generated a spike as per the input pattern. Next, from t=150 ns to t=280 ns, we rotated the pattern clockwise by 1 pixel. Had we considered only static learning, the synapses S3 and S9 would have been in the lowest conductance state and provided low current. If we then added this current to the one coming from synapses S6 and S10, the voltage generated at the MTJ input would have been 2.5 V, which would have been less than the threshold, causing the neuron not to spike. However, using the concept of STP potentiation, we increased the dynamic conductance component $G_t^{STP}$, corresponding to the synapses S3 and S9. Thus, the overall conductance also increased, and these synapses supplied a large current. When a 30% increase in $G_3$ and $G_9$ was considered, the voltage that dropped across the fixed resistor was above 3 V.

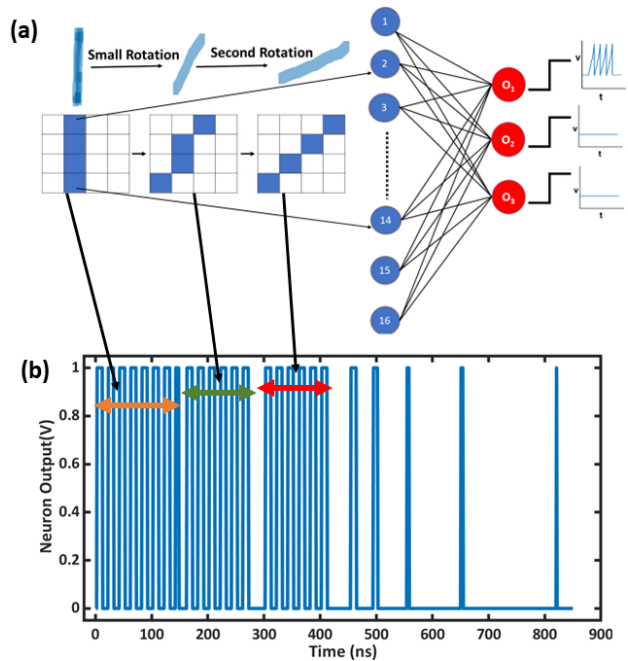

Fig. 6(a) Toy neural network with 16 input neurons and 3 output neurons for STP-based dynamic learning and inference. (b) Neuron output O1 for rotating the input digit 1. Image shows that spikes were generated for even a completely rotated pattern.

We also observed that the output neuron O1 was still able to generate spikes, as shown in the green region. Likewise, we rotated the pattern further, and with a similar approach, we clearly observed that the network was still able to recognize the digit as 1. Thus, with this toy example, we demonstrated the dynamic environment learning and inference capabilities of the proposed skyrmion device.

## VI. CONCLUSION

In this study, we proposed a skyrmion-based neuromorphic MTJ device with both long- and short-term plasticity (LTP and STP) (mixed synaptic plasticity). We proved that the plasticity could be controlled by magnetic field, SOT, and the VCMA switching mechanisms. The LTP property of the device was utilized for static image recognition on the CIFAR-10 data set and trained on the VGG-8 convolutional neural network. When an STP feature was added, the device gained additional temporal filtering ability, and the system could adapt to dynamic environment learning and show inferencing capabilities. We also demonstrated the dynamic environment learning and inference capabilities of the proposed device by using a toy example of 16 input neurons and 3 output neurons. With further advances at the algorithm level for STP-based learning, the proposed skyrmion device shows bright prospects for dynamic image recognition and clustering in more complex neuromorphic systems. Further, the skyrmions were conserved and confined to a nanotrack to minimize the skyrmion nucleation energy. The synapse device was trained and tested for emulating a deep neural network. We observed that when the skyrmion density was increased, the inference accuracy improved: 90% accuracy was achieved by the system at the highest density. The fabrication of the proposed device is in progress.

## REFERENCES

[1] A. Fert, N. Reyren, and V. Cros, "Magnetic skyrmions: Advances in physics and potential applications," *Nat. Rev. Mater.*, vol. 2, 2017, doi: 10.1038/natrevmats.2017.31.

[2] S. Woo *et al.*, "Deterministic creation and deletion of a single magnetic skyrmion observed by direct time-resolved X-ray microscopy," *Nat. Electron.*, vol. 1, no. 5, pp. 288–296, 2018, doi: 10.1038/s41928-018-0070-8.

[3] A. Bernand-Mantel, C. B. Muratov, and T. M. Simon, "Unraveling the role of dipolar versus Dzyaloshinskii-Moriya interactions in stabilizing compact magnetic skyrmions," *Phys. Rev. B*, vol. 101, no. 4, p. 45416, 2020, doi: 10.1103/PhysRevB.101.045416.

[4] S. Woo *et al.*, "Observation of room-temperature magnetic skyrmions and their current-driven dynamics in ultrathin metallic ferromagnets," vol. 15, no. February, 2016, doi: 10.1038/NMAT4593.

[5] S. Woo *et al.*, "revealed by time-resolved X-ray microscopy," *Nat. Commun.*, vol. 8, no. May, pp. 1–8, 2017, doi: 10.1038/ncomms15573.

[6] M. Ma *et al.*, " Enhancement of zero-field skyrmion density in [Pt/Co/Fe/Ir] 2 multilayers at room temperature by the first-order reversal curve ," *J. Appl. Phys.*, vol. 127, no. 22, p. 223901, 2020, doi: 10.1063/5.0004432.

[7] S. Luo and L. You, "Skyrmion devices for memory and logic applications," *APL Mater.*, vol. 9, no. 5, pp. 1–11, 2021, doi: 10.1063/5.0042917.

[8] J. Zang, M. Mostovoy, J. H. Han, and N. Nagaosa, "Dynamics of Skyrmion crystals in metallic thin films," *Phys. Rev. Lett.*, vol. 107, no. 13, pp. 1–5, 2011, doi: 10.1103/PhysRevLett.107.136804.

[9] X. S. Wang, H. Y. Yuan, and X. R. Wang, "A theory on skyrmion size," *Commun. Phys.*, vol. 1, no. 1, pp. 1–7, 2018, doi: 10.1038/s42005-018-0029-0.

[10] W. Kang, Y. Huang, X. Zhang, Y. Zhou, and W. Zhao, "Skyrmion-Electronics: An Overview and Outlook," *Proc. IEEE*, vol. 104, no. 10, pp. 2040–2061, 2016, doi: 10.1109/JPROC.2016.2591578.

[11] K. M. Song *et al.*, "Skyrmion-based artificial synapses for neuromorphic computing," *Nat. Electron.*, vol. 3, no. 3, pp. 148–155, 2020, doi: 10.1038/s41928-020-0385-0.

[12] M. Chauwin *et al.*, "Skyrmion Logic System for Large-Scale Reversible Computation," *Phys. Rev. Appl.*, vol. 12, no. 6, pp. 1–24, 2019, doi: 10.1103/PhysRevApplied.12.064053.

[13] S. Li et al., "Emerging neuromorphic computing paradigms exploring magnetic skyrmions," *Proc. IEEE Comput. Soc. Annu. Symp. VLSI, ISVLSI*, vol. 2018-July, pp. 539–544, 2018, doi: 10.1109/ISVLSI.2018.00104.

[14] G. Srinivasan, A. Sengupta, and K. Roy, "Magnetic Tunnel Junction Based Long-Term Short-Term Stochastic Synapse for a Spiking Neural Network with On-Chip STDP Learning," *Sci. Rep.*, vol. 6, no. June, pp. 1–13, 2016, doi: 10.1038/srep29545.

[15] J. Deng, V. P. K. Miriyala, Z. Zhu, X. Fong, and G. Liang, "Voltage-Controlled Spintronic Stochastic Neuron for Restricted Boltzmann Machine with Weight Sparsity," *IEEE Electron Device Lett.*, vol. 41, no. 7, pp. 1102–1105, 2020, doi: 10.1109/LED.2020.2995874.

[16] Z. He and D. Fan, "Developing All-Skyrmion Spiking Neural Network," pp. 2–4, 2017, [Online]. Available: http://arxiv.org/abs/1705.02995.

[17] Y. Huang, W. Kang, X. Zhang, Y. Zhou, and W. Zhao, "Magnetic skyrmion-based synaptic devices," *Nanotechnology*, vol. 28, no. 8, 2017, doi: 10.1088/1361-6528/aa5838.

[18] F. N. Tan et al., "High velocity domain wall propagation using voltage controlled magnetic anisotropy," *Sci. Rep.*, vol. 9, no. 1, pp. 1–6, 2019, doi: 10.1038/s41598-019-43843-x.

[19] W. G. Wang and C. L. Chien, "Voltage-induced switching in magnetic tunnel junctions with perpendicular magnetic anisotropy," *J. Phys. D. Appl. Phys.*, vol. 46, no. 8, 2013, doi: 10.1088/0022-3727/46/7/074004.

[20] P. V. Ong, N. Kioussis, P. K. Amiri, and K. L. Wang, "Electric-field-driven magnetization switching and nonlinear magnetoelasticity in Au/FeCo/MgO heterostructures," *Sci. Rep.*, vol. 6, no. June, pp. 1–8, 2016, doi: 10.1038/srep29815.

[21] C. Back, V. Cros, H. Ebert, A. Fert, M. Garst, and T. Ma, "The 2020 skyrmionics roadmap," 2020.

[22] K. Everschor-Sitte, J. Masell, R. M. Reeve, and M. Kläui, "Perspective: Magnetic skyrmions - Overview of recent progress in an active research field," *J. Appl. Phys.*, vol. 124, no. 24, 2018, doi: 10.1063/1.5048972.

[23] J. Leliaert, M. Dvornik, J. Mulkers, J. De Clercq, M. V. Milošević, and B. Van Waeyenberge, "Fast micromagnetic simulations on GPU - Recent advances made with mumax3," *J. Phys. D. Appl. Phys.*, vol. 51, no. 12, 2018, doi: 10.1088/1361-6463/aaab1c.

[24] A. Vansteenkiste, J. Leliaert, M. Dvornik, M. Helsen, F. Garcia-Sanchez, and B. Van Waeyenberge, "The design and verification of MuMax3," *AIP Adv.*, vol. 4, no. 10, 2014, doi: 10.1063/1.4899186.

[25] F. Büttner et al., "Field-free deterministic ultrafast creation of magnetic skyrmions by spin-orbit torques," *Nat. Nanotechnol.*, vol. 12, no. 11, pp. 1040–1044, 2017, doi: 10.1038/nnano.2017.178.

[26] A. Vansteenkiste, J. Leliaert, M. Dvornik, M. Helsen, F. Garcia-Sanchez, and B. Van Waeyenberge, "The design and verification of MuMax3," *AIP Adv.*, vol. 4, no. 10, pp. 0–22, 2014, doi: 10.1063/1.4899186.

[27] A. H. Lone, S. Shringi, K. Mishra, and S. Srinivasan, "Cross-Sectional Area Dependence of Tunnel Magnetoresistance, Thermal Stability, and Critical Current Density in MTJ," *IEEE Trans. Magn.*, vol. 57, no. 2, 2021, doi: 10.1109/TMAG.2020.3039682.

[28] S. Datta, *Quantum transport: Atom to transistor*, vol. 9780521631. 2005.

[29] X. Peng, S. Huang, Y. Luo, X. Sun, and S. Yu, "DNN + NeuroSim : An End-to-End Benchmarking Framework for Compute-in-Memory Accelerators with Versatile Device Technologies," pp. 771–774, 2019.